# Characterization of Left Ventricular Hypertrophy via Fractional Derivatives

Srijan Sengupta[1a], Uttam Ghosh[1b], Susmita Sarkar[1c] and Shantanu Das [2]

[1]Department of Applied Mathematics, University of Calcutta, Kolkata, India
[2]Reactor Control Systems Design Section E & I Group BARC Mumbai, India
Email:[1a]srijansengupta.math@gmail.com, [1b]uttam_math@yahoo.co.in[*], [1c]susmita62@yahoo.co.in,[2]shantanu@barc.gov.in

**Abstract**

In this paper we have used the concepts of fractional derivative of rough curves to characterize ECG of LVH patients and compared the results with normal ECGs. In mathematical language an ECG is a rough curve having Q,R,S points as non differentiable points where classical derivatives do not exist but fractional derivatives exist. We have calculated both left and right modified Riemann-Liouville fractional derivatives and their differences termed as phase transition at those non differentiable points of V1, V2, V5 and V6 leads. Investigation shows that phase transition is higher for LVH patients than normal ones. This may be a method of determination of risk factor of LVH patients before doing Echocardiogram.

## 1.0 Introduction

The fractional calculus is the generalization of ordinary differentiation and integration to an arbitrary order and is a growing field of Applied Mathematics [1-3,10]. It has many applications in Mathematical Biology, Engineering, Mathematical Economics etc [4-6]. In reality there are many functions which are everywhere continuous but non-differentiable at some or all points. The ECG, EEG, stock market curves [7,9], the coast line [7,9] curves are representation of such types of functions. ECG stands for electrocardiogram or electrocardiograph which is the pictographic representation of the electrical charge depolarization and re-polarization of the muscles of heart [11].Such ECG is crucial for the diagnosis and management of abnormal cardiac rhythms. There are several types of heart diseases that cause chest pain which may be for Left Ventricular Hypertrophy, Right Ventricular Hypertrophy etc. In this paper we have characterized the unreachable points i.e. non-differentiable points of a Normal ECG and Abnormal ECG graphs (in my case Left Ventricular Hypertrophy choosing arbitrarily) by using Jumarie definition of fractional derivative[9].It is observed that the calculated values of the phase transition at the non-differentiable points of LVH patients are higher than the values of normal ECGs.

**Left Ventricular Hypertrophy (LVH):** Left Ventricular hypertrophy (LVH) is a condition where the muscle wall of heart's left pumping chamber (ventricle) becomes enlarged and thickened (hypertrophy). This can arise in response to some reason such as high blood pressure, aortic valve stenosis, hypertrophic cardiomyopathy or a heart condition that causes the left ventricle to work harder. Left ventricular hypertrophy is very common in people who have uncontrolled high blood pressure[11].Certain genetic conditions are associated with developing hypertrophy. Women with hypertension are at higher risk of Left ventricular hypertrophy (LVH) than are men with similar blood pressure.

**Symptoms:** Left ventricular hypertrophy usually develops gradually. Patients may experience no signs or symptoms, especially during the early stages of the condition. The symptoms happen when the condition causes complications. The most common symptoms of LVH are: i. Feeling short of breath, ii. Chest pain, especially after activity, iii.Feeling dizzy or fainting, iv. Rapid heartbeat, or a pounding or fluttering sensation in chest.

**ECG of LVH (in doctor's view):**
Left ventricular hypertrophy causes a tall R wave (>25mm) in lead V5 or V6 ,a deep S wave in lead V1 or V2, inverted T wave in leads I,II,AVL,V5,V6, sometimes V4 and R waves in lead V5 or V6 plus S wave in lead V1 or V2 greater than 35 mm. **It is difficult to diagnose minor degrees of left ventricular hypertrophy from the ECG** [11]**.**

**Modified definitions of Riemann-Liouville derivative:**
The basic definitions of fractional derivatives are of Riemann-Liouville (R-L) [2], Caputo [7], and Jumarie [7,8]. To overcome the shortcoming of the R-L definition that derivative of a constant is non-zero which is contradiction of the conventional integer order calculus, Jumarie [8] first revised the R-L definition of fractional derivative in the following form.

$$D_x^\alpha f(x) = \frac{1}{\Gamma(-\alpha)} \int_0^x (x-\xi)^{-\alpha-1} f(\xi) d\xi, \text{for } \alpha < 0$$

$$= \frac{1}{\Gamma(1-\alpha)} \frac{d}{dx} \int_0^x (x-\xi)^{-\alpha} [f(\xi)-f(0)] d\xi, \text{for } 0 < \alpha < 1$$

$$= \left(f^{(\alpha-n)}(x)\right)^{(n)} \text{ for } n \leq \alpha < n+1, n \geq 1.$$

The above definition [8] is developed using left R-L derivative. Similarly modification has also been developed by us using the right R-L derivative [9]. Note in the above definition for negative fractional orders the expression is just Riemann-Liouvelli fractional integration. The modification is carried out in the R-L derivative formula, for the positive fractional orders alpha. The idea of this modification is to remove the offset value of function at start point of the fractional derivative from the function, and carry out R-L derivative as usually done for the function. [9].

**Unreachable function and Graphs**
There are many functions which are everywhere continuous but not-differentiable at some points or at all points. These functions are known as unreachable functions[8]. The function (i) $f(x) = |x|$ is unreachable at the point $x = 0$. This means they have no integer order derivative at $x = 0$ but have ½ order derivative at that point [1]. Unreachable graphs are diagrammatic representation of unreachable functions. The ECG graphs are such types of unreachable graphs. They have unreachable points Q, R, S in the 'PQRST' wave at which classical derivatives do not exist but fractional derivatives exist.

**Phase Transition**
We define the phase transition at a non differentiable point of a continuous graph as the difference between the left and right modified R-L derivatives at that point.

The organization of the paper is as follows: in section 2.0 some fractional calculus technique to characterize Left Ventricular Hypertrophy are given, in section 3.0 we use the concept of fractional derivative in normal and abnormal ECG curve ; and finally this paper is concluded in section 4.0 .

**2.0 Fractional Calculus Technique for Characterizing Left Ventricular Hypertrophy:**
In this section we shall describe the fractional calculus technique for characterizing left ventricular hypertrophy from ECG diagrams. This mathematical technique is used to calculate both left and right fractional derivatives and hence the phase transitions at the Q,R,S points of the QRS complexes of the V1, V2, V5 and V6 leads of ECG graphs. For this reason we construct theorems for characterizing ECGs.
**Theorem:** Letus consider the function
$$f(x) = \begin{cases} ax+b, & p \leq x \leq q \\ cx+d, & q \leq x \leq r \end{cases}$$
with $a \neq c$. It is continuous at x=q such that aq+b=cq+d but not differentiable at that point. Then left fractional derivative, right fractional derivative and phase transition at that point x=q are respectively:

$$f_L^{(\alpha)}(q) = \frac{a(q-p)^{1-\alpha}}{\Gamma(2-\alpha)}, \quad f_R^{(\alpha)}(q) = \frac{c(r-q)^{1-\alpha}}{\Gamma(2-\alpha)};$$

$$P.T = \frac{\left(a(q-p)^{1-\alpha} - c(r-q)^{1-\alpha}\right)}{\Gamma(2-\alpha)}$$

In the above theorem we consider a function which is linear in both sides of non-differentiable point x=q. Similar type of theorems are constructed for both sides nonlinear and one side linear & another side nonlinear functions to analyze the ECG graphs. All these cases have to be used in this paper in next section.

**3.0 Applications of fractional derivative in ECG Graphs**
Now we shall characterize the ECG graphs by the help of fractional derivative and compare normal ECGs with abnormal ECGs (LVH). For this purpose we shall consider ½- order fractional derivatives. The non-differentiable points Q, R, S of QRS complexes of PQRST wave of any leads are used here usually points, not as wave. Also we shall calculate the fractal dimension of the considered leads of ECG graphs. If Q or S point is not prominent at QRS complex of any lead of the ECGs under consideration then we cannot find the Left Fractional Derivative and Right Fractional Derivative at that point. We have denoted those cases by 'NA' i.e. 'Not Arise'. To investigate the characteristics of the ECG we here consider normal ECG and problematic ECG(in our case LVH ECG to be compared with normal ECG).
Since LVH is characterized by deep S wave in V1 and V2 leads and long R wave in V5 and V6 leads.So our concern to find any distinguishing measurements of P.T values at non-differentiable points on S wave in V1 and V2 and R wave in V5 and V6 leads to characterize the problematic ECG (in our case LVH) with normal ECG.
Thus our paper contributed only P.T. values at non-differentiable points at those leads for LVH ECGs. So our concern to find any distinguishing measurements of P.T values at non-differentiable points on S wave in V1 and V2 and R wave in V5 and V6 leads to characterize the problematic ECG (in our case LVH) with normal ECG.Following tables are new and constructed from our fractional calculus methodology.

## Characterization ECG

The considerable methodology which is now used for normal ECG as well as for problematic ECG (LVH)[12] also shown to help us a strong comparison between Normal ECGs and LVH ECGs in this section below.

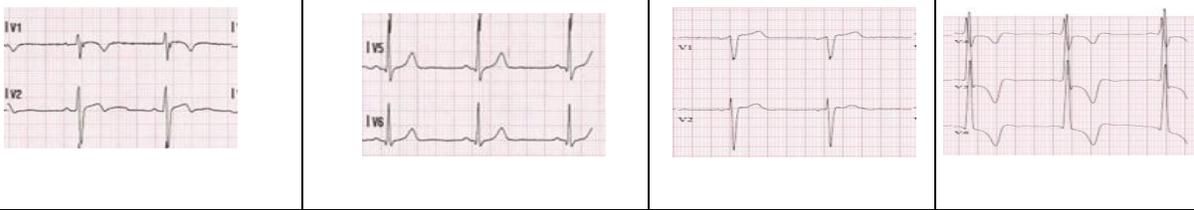

| Normal ECG sample | | | | | Problematic ECG sample | | | |
|---|---|---|---|---|---|---|---|---|
| PQRST waves | A | B | C | D | PQRST waves | A | B | C | D |
| 1st | 9.5 | 21.5 | 18 | 14.5 | 1st | 12 | 27 | 56 | 56.5 |
| 2nd | 10.5 | 23 | 18 | 14.5 | 2nd | 11 | 25 | 54.1 | 56 |
| 3rd | | | 18 | 13.5 | 3rd | | | 59 | 53.2 |
| E=A+C | F=A+D | G=B+C | H=B+D | | E=A+C | F=A+D | G=B+C | H=B+D | |
| 27.5 | 24 | 39.5 | 36 | | 68 | 68.5 | 83 | 83.5 | |
| 27.5 | 24 | 39.5 | 36 | | 66.1 | 68 | 81.1 | 83 | |
| 27.5 | 23 | 39.5 | 35 | | 71 | 65.2 | 86 | 80.2 | |
| 28.5 | 25 | 41 | 37.5 | | 67 | 67.5 | 81 | 81.5 | |
| 28.5 | 25 | 41 | 37.5 | | 65.1 | 67 | 79.1 | 81 | |
| 28.5 | 24 | 41 | 36.5 | | 70 | 64.2 | 84 | 78.2 | |

**Table 1A:** Length of different parts of ECG: A= Length of S wave (mm) in V1 lead, B= Length of S wave (mm) in V2 lead, C= Length of R wave (mm) in V5 lead, D=Length of R wave (mm) in V6 lead respectively.

Here, we see that in table-1A lengths of S and R wave in V1,V2 and V5,V6 respectively are all greater than 25 mm except length of S wave in V1 lead for problematic ECG , these measurement is below 25mm for normal ECG and length of E,F,G and H are all greater than 35 mm of that table for problematic ECG, these measurement is also below for normal ECG. So from Doctor's point of view this patient with problematic ECG has cardiac problem which called Left Ventricular Hypertrophy.

Now we have to calculate above mentioned techniques to characterize ECGs for comparing problematic ECG with normal ECGgiven below.

| | | P.T. values of V1 | P.T. values of V2 | P.T. values of V5 | P.T. values of V6 |
|---|---|---|---|---|---|
| 1st | For PQ:QR at Q | 6.383076 | NA | 30.244018 | 25.489442 |
| | For QR:RS at R | 21.542883 | 34.4156 | 62.234996 | 40.064340 |
| | For RS:ST at S | 25.532306 | 44.2073 | 36.896290 | 20.117226 |
| 2nd | For PQ:QR at Q | 5.077706 | 10.1554 | 30.319613 | 21.243420 |
| | For QR:RS at R | 21.833282 | 46.8581 | 52.419806 | 42.829167 |
| | For RS:ST at S | 24.734422 | 57.4477 | 28.483269 | 25.239477 |
| 3rd | For PQ:QR at Q | | | 31.915382 | 23.823144 |
| | For QR:RS at R | | | 61.437111 | 37.340191 |
| | For RS:ST at S | | | 35.534215 | 18.190962 |

**Table 1B:** Phase transition at the non-differentiable points Q,R,S of V1,V2,V5 and V6 leads of Normal ECG sample

| | | P.T. values of V1 | P.T. values of V2 | P.T. values of V5 | P.T. values of V6 |
|---|---|---|---|---|---|
| 1st | For PQ:QR at Q | NA | 15.1388 | 62.1737 | 62.5706 |
| | For QR:RS at R | 15.2332 | 49.2011 | 141.7848 | 118.2342 |
| | For RS:ST at S | 18.0541 | 40.8057 | 81.8757 | NA |
| 2nd | For PQ:QR at Q | NA | 12.3608 | 61.1582 | 60.0624 |
| | For QR:RS at R | 15.1027 | 39.2575 | 135.0026 | 115.2309 |
| | For RS:ST at S | 16.7741 | 36.9892 | 81.0753 | NA |

| | | | | |
|---|---|---|---|---|
| 3rd | For PQ:QR at Q | | 62.2868 | 55.1727 |
| | For QR:RS at R | | 142.2155 | 185.3884 |
| | For RS:ST at S | | 84.8053 | NA |

**Table 1C:** Phase transition at the non-differentiable points Q,R,S of V1,V2,V5 and V6 leads of Problematic ECG(LVH) sample

## 4. Discussion

In this paper we have studied characteristics of normal ECG graphs and ECG graphs of LVH patients, by finding fractional derivatives at non-differentiable points. From the above tables it is observed that Phase Transition (P.T) values are maximum at the point R and S of QRS complexes at V1, V2 and V5, V6 leads of problematic ECG respectively. We have recorded and compared the P.T values at the point R and S of V1, V2 and V5, V6 leads of different ECG leads. The P.T. values of different non-differentiable points of the V1, V2, V5, V6 leads in normal ECG are less than 50. For values 50 to 65 the patients are prone to LVH where as patients having P.T. values above 65 in the V1, V2, V5, V6 leads are suffering from LVH problem. But these ranges are not present in many cases such as if there are any other problem together with LVH; also if the nature of any lead behaves like a normal ECG lead (since ECG do not found any disease at first or minor stage in ECG tracing paper many times). In those case Doctors will decide to do another tests like ECO, TMT, Angiography of the patient to detect exact problem of that patient. The values of P.T. of normal ECG leads for different non-differentiable points are low but it increases abruptly for LVH patients. From our samples we can conclude that if the phase transition value is greater than 50 then the person will be in danger zone. Thus by studying large number of ECG it is possible to find the a suitable range for the fractal dimension of the ECG leads and phase transition (P.T) values at the non-differentiable points that will help the doctors to determine the LVH conditions of patients. We will report other ailments in our next study. However, this type of study is not reported elsewhere. This method is a new method we are reporting for the first time-could be an aid for differential diagnostics in medical science.

**Compliance with Ethical Standards**
Authors declare that none of them have any conflict of interest.

**Ethical approval**
All ECG have been downloaded from internet

## 5.0 Acknowledgement
Authors thank **Dr. TridipSengupta**, Ex-Senior Medical Officer Cardiology, R. G. Kar Medical College & Hospital, Kolkata, and **Dr. ManoranjanMandal**, Department of Cardiology, N. R. S. Medical College, Kolkata, for their valuable guidance in understanding ECG graphs from medical point of view, and encouragement on this new idea to have a characterization studies for ECG.